\documentclass[aps,prx,twocolumn,superscriptaddress]{revtex4-2}
\usepackage[utf8]{inputenc}
\usepackage[english]{babel}
\usepackage{graphicx}
\usepackage{chemformula}
\let\ce\ch
\usepackage{siunitx}
\usepackage{color}
\usepackage{amsmath,amssymb}
\usepackage{tabularx}
\usepackage{multirow}
\usepackage{changes}

\newcommand*{\scol}{3.375in}
\newcommand*{\sdcol}{4.875in}

\newcommand*{\sect}{Sec.\,}

\newcommand*{\fig}{Fig.\,}
\newcommand*{\eq}[1]{Eq.\,(#1)}
\newcommand*{\tab}{Tab.\,}
\newcommand*{\reftaken}{Ref.\,}

\newcommand*{\supp}{Supp.\,}
\newcommand*{\etal}{\emph{et~al. }}

\newcommand*{\ie}{i.e.\,}
\newcommand*{\eg}{e.g.\,}
\newcommand*{\cf}{cf.\,}
\renewcommand{\vec}[1]{\boldsymbol{#1}}

\newcommand*{\sderr}[3]{$#1(#2)$\,\si{#3}}

\begin{document}
\title{Microscopics of Quantum Annealing in the Disordered\\Dipolar Ising Ferromagnet \ce{LiHo_{x}Y_{1-x}F4}}

\author{S.~S{\"a}ubert}
\email[]{steffen.saeubert@colostate.edu}
\affiliation{Department of Physics, Colorado State University, Fort Collins, Colorado 80523-1875, USA}

\author{C.L.~Sarkis}
\affiliation{Department of Physics, Colorado State University, Fort Collins, Colorado 80523-1875, USA}

\author{F.~Ye}
\affiliation{Neutron Scattering Division, Oak Ridge National Laboratory, Oak Ridge, Tennessee 37831, USA}

\author{G.~Luke}
\affiliation{Department of Physics and Astronomy, McMaster University, Hamilton, Ontario L8S 4M1, Canada}
\affiliation{Brockhouse Institute for Materials Research, McMaster University, Hamilton, Ontario L8S 4M1, Canada}
\affiliation{TRIUMF, Vancouver, British Columbia V6T 2A3, Canada}

\author{K.A.~Ross}
\email[]{kate.ross@colostate.edu}
\affiliation{Department of Physics, Colorado State University, Fort Collins, Colorado 80523-1875, USA}
\affiliation{Quantum Materials Program, CIFAR, MaRS Centre, West Tower 661 University Ave., Suite 505, Toronto, ON, M5G 1M1, Canada}

\date{\today}

\begin{abstract}
Quantum annealing (QA) refers to an optimization process that uses quantum fluctuations to find the global minimum of a rugged energy landscape with many local minima.  Conceptually, QA is often framed in the context of the disordered transverse field Ising model, in which a magnetic field applied perpendicular to the Ising axis tunes the quantum fluctuations and enables the system to tunnel through energy barriers, and hence reach the ground state more quickly. A solid state material closely related to this model, \ce{LiHo_{0.45}Y_{0.55}F4}, was shown to exhibit faster dynamics after a QA protocol, compared to thermal annealing (TA), but little is known about the actual process of optimization involved or the nature of the spin correlations in the state that is reached.  Here, we report on the microscopics of QA in this material using diffuse magnetic neutron scattering. 
Comparing a QA to a TA protocol that reach the same end-point, we find very similar final diffuse scattering which consists of pinch-point scattering, largely consistent with critical scattering near a phase boundary in the dipolar Ising ferromagnetic model.  However, comparing the time evolution at the end of the protocols, we find that the spin correlations evolve more significantly after TA, suggesting that QA produces a state closer to equilibrium. We also observe experimental evidence that the transverse field produces random fields, which had been previously predicted for this material and studied in other contexts. Thus, while the material does exhibit a “quantum speedup” under quantum annealing conditions, it is not a simple annealing problem; the energy landscape being optimized is changing as the optimization proceeds. 
\end{abstract}

\clearpage

\pacs{}

\maketitle

\section{Introduction}
\label{sec:introduction}

Quantum annealing (QA) refers to a method for solving optimization problems through quantum fluctuations, rather than thermal fluctuations used in thermal annealing (TA).  QA has been argued to reach the ground state faster than TA,  \ie it produces a ``quantum speedup'', for some complex optimization problems \cite{1994_Finnila_ChemicalPhysicsLetters,2015_Dutta_Book}. 
These problems include, for instance, the 
traveling salesman problem that has a many-valleyed cost landscape \cite{2008_Das_RevModPhys}, or, in physics, finding the ground state of a Hamiltonian with a complex energy landscape (\eg spin-glasses). When energy barriers separating local minima are very high, finding the global minimum via TA requires an infinitely slow annealing protocol, showing that a more efficient method is required in these cases \cite{1983_Kirkpatrick_Science,1984_Geman_IEEETransPatternAnalMachIntell,2015_Dutta_Book}. A TA protocol involves introducing tunable noise to help the system escape local minima, which in a materials context is equivalent to tuning thermal fluctuations via the temperature.  Such protocols start at high temperature and slowly reduce it, eventually settling into what one hopes is the global minimum.  Meanwhile, in QA, a parameter is introduced which controls the level of quantum fluctuations, allowing the system to tunnel through high (but sufficiently narrow) energy barriers rather than thermally jumping over them \cite{1983_Kirkpatrick_Science,1989_Ray_PhysRevB}, and the quantum fluctuations are slowly reduced to eventually reach the ground state. Whether QA really works better than TA has been an active topic of study. A faster optimization of the ground state energy using QA has been observed for some problems \cite{2002_Santoro_Science,2004_Martonak_PhysRevE}, but not for others \cite{2005_Battaglia_PhysRevE,2009_Matsuda_NewJPhys}.
Due to the successes of QA in optimization of some complex energy landscape problems, this technique is used in ``adiabatic'' quantum computers, such as the D-wave system \cite{2011_Johnson_Nature,2014_Boixo_NatPhys}, which, incidentally, is itself becoming a tool used successfully in studying geometrically frustrated Ising systems \cite{2020_Kairys_PRXQuantum,2021_King_NatCommun}.

A model used to study QA is the transverse field Ising model (TFIM) \cite{1998_Kadowaki_PhysRevE}, which is experimentally realized in the tetragonal material \ce{LiHoF4} and its disordered variant \ce{LiHo_{x}Y_{1-x}F4} \cite{2011_Gingras_JPhysConfSer}. \ce{LiHoF4} is a dipolar coupled Ising ferromagnet where the crystal electric field environment leads to a strong Ising anisotropy of the \ce{Ho^{3+}} ions along the crystallographic $c$-direction \cite{2011_Gingras_JPhysConfSer}. The marginal dimension of the dipolar Ising ferromagnet separating mean-field from non-mean-field behavior is $d^{\star}\,=\,3$ and thus mean-field theory describes the behavior of the pure compound quantitatively \cite{1973_Aharony_PhysRevB}, although the hyperfine interaction between the electronic spins and the nuclear spin $7/2$ of holmium needs to be taken into account a low temperatures \cite{1984_Mennenga_JournalofMagnetismandMagneticMaterials,1996_Bitko_PhysRevLett,2005_Ronnow_Science,2005_Schechter_PhysRevLett}. The Hamiltonian describing the resulting version of the TFIM is given by
\begin{align}
    \mathcal{H}\,=\,\sum_{(i,j)} J_{ij} S_{i}^{z} S_{j}^{z} - H_{\perp} \sum_{i} S_{i}^{x}
\end{align}
where the sum is over all pairs (not just nearest neighbors), $J_{ij}$ is the exchange interaction between Ising spins $i$ and $j$ with the Ising axis along $z$, and $H_{\perp}$ is a magnetic field applied along $x$ (transverse to the Ising direction). 
The transverse field $H_{\perp}$ acts as a tunneling term for QA, which destroys the Ising order for $H_{\perp}\,>\,H_c$ even at zero temperature. Due to the disruption of the dipolar field symmetries, diluting the system with non-magnetic \ce{Y^{3+}} produces interaction disorder and, as we shall emphasize below, random longitudinal fields correlated with the interaction disorder \cite{2006_Schechter_PhysRevLett,2006_Tabei_PhysRevLett,2007_Silevitch_Nature,2007_Schechter_JPhysCondensMatter,2008_Schechter_PhysRevB,2008_Tabei_PhysRevB}. The transverse-field-induced random fields have been shown to cause a deviation from MFT of the phase boundaries \cite{1997_Bitko_PhD,2007_Silevitch_Nature, 2016_Babkevich_PhysRevB} and the critical behavior \cite{2007_Silevitch_Nature}. The frustration and disorder in the diluted material leads to a spin glass regime for low $x$ compositions \cite{2008_Ancona-Torres_PhysRevLett,2011_Gingras_JPhysConfSer}, although for $x\,\leq\,0.25$ the existence of a spin glass (rather than an ``anti-glass'') is debated \cite{2008_Ancona-Torres_PhysRevLett,2007_Jonsson_PhysRevLett,2010_Jonsson_ArXiv08031357Cond-Mat}.  
The full microscopic Hamiltonian describing the diluted dipolar coupled TFIM, including random field effects, can be found in \reftaken\cite{2006_Tabei_PhysRevLett}.

\begin{figure}[tb]
	\centering
	\includegraphics[width=\scol]{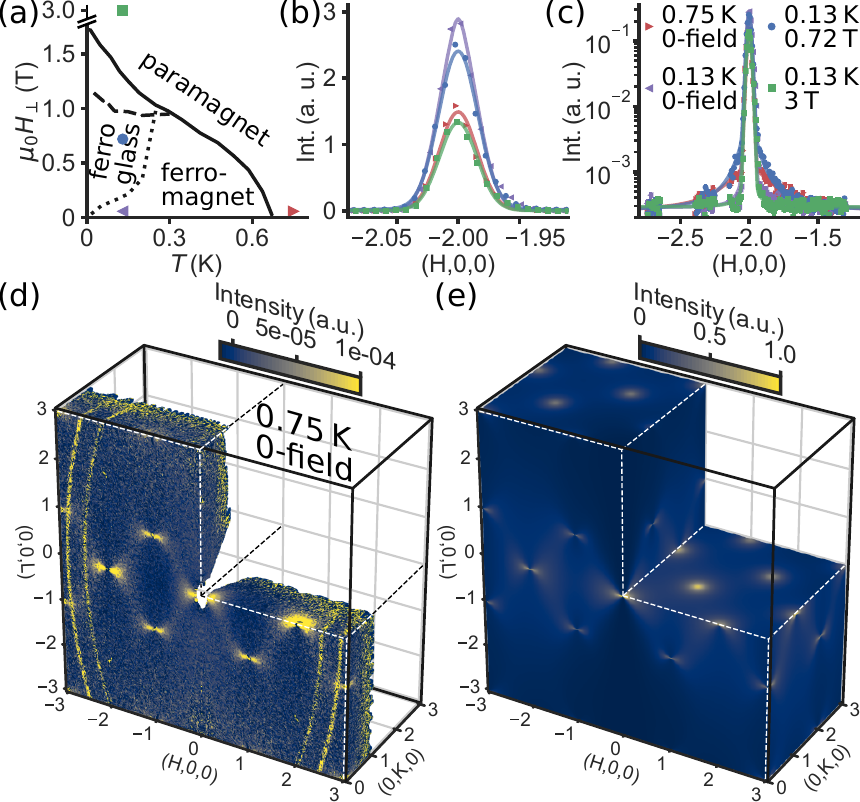}
	\caption{
	Diffuse magnetic neutron scattering in \ce{LiHo_{x}Y_{1-x}F4} ($x\,=\,0.45$). 
	(a) Transverse field vs. temperature phase diagram of \ce{LiHo_{x}Y_{1-x}F4} ($x\,=\,0.44$) showing the para- and ferromagnetic phases, as well as the ferroglass phase at low temperatures and finite fields. The phase diagram is taken from \reftaken\cite{1997_Bitko_PhD,1999_Brooke_Science}. 
	Colored symbols represent positions in the phase diagram corresponding to data shown in panels (b) and (c).  
	(b) Cuts through the $(-2,0,0)$-Bragg peak along $H$ ($L\,=\,[-0.01,0.01]$ and $K\,=\,[-0.01,0.01]$) at different temperatures and fields in the paramagnet in zero-field (red), ferromagnet in zero-field (purple), ferroglass (blue), and field-polarized paramagnet (green). Solid lines are fits to the data described in the main text. (c) Same data as in (b) on a logarithmic intensity scale to emphasize the diffuse contribution.
	(d) Background-subtracted experimental data taken at \SI{0.75}{K} in zero-field. Diffuse scattering emerges at magnetic Bragg peaks. Notice for example the $(-2,0,2)$-Bragg peak where only nuclear Bragg intensity is observed, and no diffuse scattering. The diffuse scattering emerges as pinch-point scattering in the $\mathit{HL}$- and $\mathit{KL}$-planes, which looks like diffuse spots in the $\mathit{HK}$-planes.
	(e) Mean-field simulation calculated from the parameters obtained from fitting \eq{\ref{eq:convofit}} to the experimental data near $(-2,0,0)$.
	}
	\label{fig:fig1_pd_lcuts_3d}
\end{figure}

The combination of a complex energy landscape (spin glass), and the ability to tune quantum fluctuations using the transverse field suggests that \ce{LiHo_{x}Y_{1-x}F4} could be an ideal platform for studying QA. Indeed, this has been explored already for one particular holmium concentration, $x=0.44$ \cite{1999_Brooke_Science}. For this composition, the disorder manifests as a ``ferroglass'' phase at low temperatures and finite transverse field that interrupts the regular ferromagnetic state, \cf the phase diagram in \fig\ref{fig:fig1_pd_lcuts_3d}(a) \cite{1999_Brooke_Science,2011_Gingras_JPhysConfSer}.  This state displays a net magnetization, but slower dynamics than the FM phase, and has been characterized as ``critical'' due to a logarithmic dependence of the susceptibility on frequency \cite{1999_Brooke_Science}. 

In the pioneering QA study of the $x=0.44$ composition, performed by Brooke \etal \cite{1999_Brooke_Science}, a difference between QA and TA was observed in the ferroglass phase. The dynamics, observed by the frequency dependence of the ac susceptibility, were found to be faster after QA. This was speculated to be a result of having reached the ground state with QA, but not TA, though this was not directly shown. A theoretical study did show the ``quantum speedup'' (faster optimization) directly by studying the 2D random Ising model (a prototype spin-glass) in a transverse field using classical and quantum Monte Carlo methods \cite{2002_Santoro_Science}. 
However, these two results are not identical; in the experimental case, the dynamics of the resulting state were observed to be faster after QA, but it is not clear whether it is truly the ``ground state''.  Meanwhile, in the theoretical work, only the time taken to get to the ground state was considered, not the resulting dynamics, and the model is greatly simplified compared to the model believed to be appropriate for \ce{LiHo_{x}Y_{1-x}F4}. There are therefore some fundamental unanswered questions about QA in this material, including how the optimization process occurs, and what the nature of the resulting microscopic spin correlations is after QA compared to TA.
To address these questions, we studied the microscopics of QA in \ce{LiHo_{x}Y_{1-x}F4} using diffuse magnetic neutron scattering in three dimensions, performing the same protocols as were initially used to demonstrate QA in this material. 
The results presented in this article show that QA in \ce{LiHo_{x}Y_{1-x}F4} does achieve a state that is closer to the equilibrium ground state, compared to TA. This is made clear by the time-dependence of the diffuse magnetic scattering after reaching the end-point. Additionally, we also clearly observe the effects of the random fields induced by the transverse field, which we argue should have significant implications for the process of QA in \ce{LiHo_{x}Y_{1-x}F4}.

\section{Materials and Methods}
\label{sec:materials_and_methods}

\subsection{Neutron scattering experiments}
\label{sub:neutron_scattering_experiments}

Neutron scattering experiments were performed at the Spallation Neutron Source (SNS) at Oak Ridge National Laboratory (ORNL), Oak Ridge, TN, USA. We used the statistical chopper spectrometer CORELLI \cite{2008_Rosenkranz_Pramana-JPhys} to study how the real-space spin correlations change under quantum and thermal annealing using diffuse magnetic neutron scattering. CORELLI allows the measurement of large volumes of reciprocal space of diffuse scattering by combining white-beam time-of-flight Laue diffraction with an energy discrimination through a cross-correlation method \cite{2008_Rosenkranz_Pramana-JPhys,2018_Ye_JApplCryst}. Experiments were performed on a \SI{3.8}{g} high-quality \ce{LiHo_{x}Y_{1-x}F4} ($x\,=\,0.45$) single crystalline sample of cylindrical shape. The sample, grown via the Bridgeman method, was purchased from TYDEX J.S.Co. (St. Petersburg) and was previously studied in \reftaken\cite{2009_Rodriguez_PhD,2010_Quilliam_PhD}. During our experiment, the sample was oriented carefully (within \SI{1}{\degree}, \cf\supp\ref{sup:sample_and_sample_orientation}, \fig\ref{fig:fig_sup_samplealignment}) with the crystallographic $b$-axis---which is transverse to the Ising axis---along the field direction, perpendicular to the crystallographic $a$-axis and the Ising direction $c$. Thus, due to the vertical magnetic field geometry used, the $(H,0,L)$ plane is the horizontal scattering plane. The sample was rigidly fixed in an oxygen-free copper sample holder and firmly bolted to the mixing chamber of a dilution refrigerator, providing excellent thermal contact, and reaching a base temperature of \SI{0.13}{K}, \cf\supp\ref{sup:sample_and_sample_orientation}. The sample was rotated in steps of \SI{3}{\degree} for a total range of \SI{102}{\degree}. Using an incident white beam thus covers a large volume in reciprocal space that is limited in $K$ to $\pm-1$, since $K$ is vertical and the detectors in that direction are partially shielded by the magnet sample environment. During the annealing protocols, the sample rotation was fixed so that the $(-2,0,0)$-peak intensity was maximized, thus a single scan covers a volume in reciprocal space including the $(-2,0,0)$-peak and the diffuse scattering emerging from it. The quasi-elastic signal was extracted using the cross-correlation method \cite{2008_Rosenkranz_Pramana-JPhys}. The energy resolution of CORELLI depends on the incident neutron energy $E_{\mathrm{inc}}$ and is $\frac{\Delta E}{E_{\mathrm{inc}}}\,\sim\,2\mathrm{-}8\,\si{\percent}$, getting worse for higher energy neutrons \cite{2018_Ye_JApplCryst}. Thus, at small momentum transfer $Q$ the cross-correlation method allows the reconstruction of the quasi-elastic signal with high energy resolution which gets worse at higher $Q$ regions. It should be noted here, that, with an energy range of $E_{\mathrm{inc}}\,=\,10\mathrm{-}200\,\si{meV}$ we also integrate over the first excited crystal electric field (CEF) state at about \SI{0.77}{meV} \cite{1975_Hansen_PhysRevB}, at least for some of the $Q$ range.  However, the CEF level is expected to be relatively featureless in $Q$, aside from the magnetic form factor. Data were collected for \SI{3}{pcm} (\SI{1}{pcm} is equivalent to \SI{8.3e10}{C}) per rotation step during the large volume scans, and \SI{5}{pcm} per temperature/field step during the annealing protocols. If not stated otherwise, a background was subtracted from the data shown in this manuscript by using a field-polarized dataset at low temperatures, \ie taken at \SI{0.13}{K} and \SI{3}{T}.

\subsection{Annealing protocols}
\label{sub:annealing_protocols}

We compare quantum versus thermal annealing after carefully running two transverse field-temperature protocols from the same start-point to the same end-point. These protocols were designed to follow exactly the annealing protocols used by Brooke \etal \cite{1999_Brooke_Science}. The annealing protocols---discussed in \sect\ref{sub:quantum_vs_thermal_annealing}---are shown in \fig\ref{fig:fig2_pd_prots_lcuts} (transverse field vs. temperature phase diagram and protocols in (a), temperature and field evolution as a function of annealing time in (b) and (c), respectively). Both protocols start at the same location at \SI{0.75}{K} in zero-field. They also end at the same location in the ferroglass at \SI{0.13}{K} and \SI{0.72}{T}, after a total annealing time for each protocol of \SI{16.5}{h}. The end-point in the ferroglass is reached two ways depending on the annealing protocol. i) During the QA protocol first a field of \SI{2.4}{T} was applied and the sample was then cooled from \SI{0.75}{K} to the base temperature of \SI{0.13}{K}. Finally, during the annealing part of this protocol, the field was slowly removed to reach the end-point at \SI{0.13}{K} and \SI{0.72}{T}. ii) The TA protocol, on the other hand, starts with the annealing by slowly cooling the sample to base temperature in zero-field. Finally, field was applied to enter the ferroglass and reach the same end-point.

\section{Results and Discussion}
\label{sec:results_and_discussion}

\subsection{Diffuse Magnetic Neutron Scattering}
\label{sub:diffuse_magnetic_neutron_scattering}

The evolution of the magnetic Bragg and diffuse scattering throughout the phase diagram of \ce{LiHo_{x}Y_{1-x}F4} is shown in \fig\ref{fig:fig1_pd_lcuts_3d}. Data shown were obtained at different locations in the phase diagram as shown in (a), \ie in the paramagnet just above the phase transition in zero-field, ferromagnet at zero-field, ferroglass in an applied transverse field, and in the field-polarized paramagnet. We compare the data using an intensity vs. $H$ cut through the $(-2,0,0)$-Bragg peak shown on a linear (b) and logarithmic (c) intensity scale. Outside the ordered phases, above the phase transition in zero-field and in the field-polarized state, we observe only the nuclear contribution to the Bragg intensity. Although the field-polarized state should have a net moment along the field direction, which in principle would result in a magnetic Bragg peak at $(-2,0,0)$, the induced moment is likely to be very small given the single-ion Ising anisotropy in zero field \cite{2008_Tabei_PhysRevB,2008_Tabei_PhysRevBa}. Inside the ordered phase the magnetic contribution increases the Bragg intensity, with the zero-field data (ferromagnet) showing stronger intensity as in finite field (ferroglass). The diffuse magnetic scattering takes a Lorentzian shape and can be seen in (c). Strong diffuse intensity is observed at \SI{0.75}{K} in zero-field (in the paramagnet just above the phase transition) and  at \SI{0.13}{K} at \SI{0.72}{T} (in the ferroglass just below the phase boundary) indicating that the observed diffuse scattering comes from critical scattering at the phase transition.  Indeed, an earlier neutron scattering study on a series of \ce{LiHo_{x}Y_{1-x}F4} samples, including $x=0.46$, showed the presence of diffuse critical scattering near the phase boundaries \cite{2016_Babkevich_PhysRevB}.

To understand these data, a more detailed discussion of the magnetic phase diagram of \ce{LiHo_{x}Y_{1-x}F4} is necessary. The phase diagram shown in \fig\ref{fig:fig1_pd_lcuts_3d}(c) was obtained via AC susceptometry by \reftaken\cite{1997_Bitko_PhD,1999_Brooke_Science} for \ce{LiHo_{x}Y_{1-x}F4} with $x\,=\,0.44$, instead of $x\,=\,0.45$ as for our sample. The solid line phase boundary was obtained by tracking the peak in $\chi^{\prime}_{\mathrm{ac}}$, while the dashed line was obtained by tracking the peak in $\chi^{\prime\prime}_{\mathrm{ac}}$. The dotted line separating ferromagnet from ferroglass was obtained via a crossover to a logarithmically divergent $\chi^{\prime}_{\mathrm{ac}}$. 
During our neutron scattering experiment we integrate over both, $\chi^{\prime}$ and $\chi^{\prime\prime}$, thus expect a phase boundary that is between these two cases.
From the dependence of the $(-2,0,0)$ peak intensity, which we tracked throughout the QA and TA protocols (discussed in detail in \sect\ref{sub:quantum_vs_thermal_annealing}) we find $T_{\mathrm{C}}\,=\,$\sderr{0.65}{1}{K}  in zero-field, in agreement with literature for $x=0.44$ (\sderr{0.67}{1}{K} \cite{1999_Brooke_Science}). From the QA protocol, on the other hand, we find $H_{\mathrm{C}}\,=\,$\sderr{0.80}{4}{T} at \SI{0.13}{K}, \ie the transition from the field polarized state to the ferroglass happens at a significantly lower field than reported in literature (obtained from the data in \reftaken\cite{1997_Bitko_PhD,1999_Brooke_Science}:  $H_{\mathrm{C},\chi^{\prime}}\,=\,$\sderr{1.30}{5}{T} and $H_{\mathrm{C},\chi^{\prime\prime}}\,=\,$\sderr{0.98}{5}{T} at \SI{0.13}{K}). The deviation from the critical fields and temperatures reported in literature can be explained by two main influences: i) Neutron scattering integrates over $\chi^{\prime}$ and $\chi^{\prime\prime}$. ii) The small deviation of the magnetic field from the ideal transverse direction (\SI{1.15}{\degree}). Previous work has shown---for the pure compound \ce{LiHoF4}---that a misalignment of the Ising axis to the applied field of only \SI{1}{\degree} reduces the critical field by $\sim$~\SI{10}{\percent} \cite{2018_Rucker_PhD}. Taking all this into account, the end-point at \SI{0.13}{K} and \SI{0.72}{T}, where a lot of our study focuses on, is in the ferroglass state of \ce{LiHo_{x}Y_{1-x}F4} ($x\,=\,0.45$), but very close to the phase transition.  Thus, the diffuse scattering observed at the end-point (Fig. \ref{fig:fig1_pd_lcuts_3d}(c)) is likely attributable to critical scattering.

\fig\ref{fig:fig1_pd_lcuts_3d}(d) shows the large reciprocal space we obtained during the volume scans (the intensity vs. $H$ cuts in (b) and (c) were extracted from similar scans). Data shown were obtained at \SI{0.75}{K} in zero-field and corrected for background by subtracting a field-polarized dataset. The diffuse scattering emerges as pinch-point scattering at magnetic Bragg-peak positions, as was previously observed in the sister compounds \ce{LiHo_{x}Er_{1-x}F4} \cite{2013_Piatek_PhysRevB,2014_Piatek_PhysRevB} and \ce{LiTbF4} \cite{1976_Als-Nielsen_PhysRevLett}.  We note that pinch-points are famous in the context of spin ice, which exhibits dipolar spin correlations due to an emergent coulomb phase \cite{2003_Moessner_PhysRevB,2007_Fennell_NatPhys,2009_Fennell_Science,2020_Twengstrom_PhysRevResearch}.  However, in the case of \ce{LiHo_{x}Y_{1-x}F4}, the presence of pinch-points does not indicate exotic physics since the dipolar correlations result directly from the dipole-dipole interactions. In the $\mathit{HL}$- and $\mathit{KL}$-planes the pinch-points take the form of ``bow-ties'' that extend throughout each Brillouin zone for which the magnetic structure factor is non-zero.   In the $\mathit{HK}$-planes, the diffuse scattering of the pinch-points look like diffuse spots. The intense rings of scattering seen in \fig\ref{fig:fig1_pd_lcuts_3d}(d) are due to powder diffraction from the copper sample holder that, due to the large intensities, cannot be perfectly subtracted with a background correction.

The diffuse scattering can be explained by the effects of dipolar coupled Ising moments. This will be discussed in detail in \sect\ref{sub:random_field_effects}. For the remainder of this paper we focus on the diffuse scattering in the $(H,0,L)$-plane, as this plane allows to study magnetic scattering along the Ising direction ($L$) and perpendicular to it ($H$).

\subsection{Quantum vs. thermal annealing} 
\label{sub:quantum_vs_thermal_annealing}

\begin{figure*}[hbtp]
	\centering
	\includegraphics[width=\sdcol]{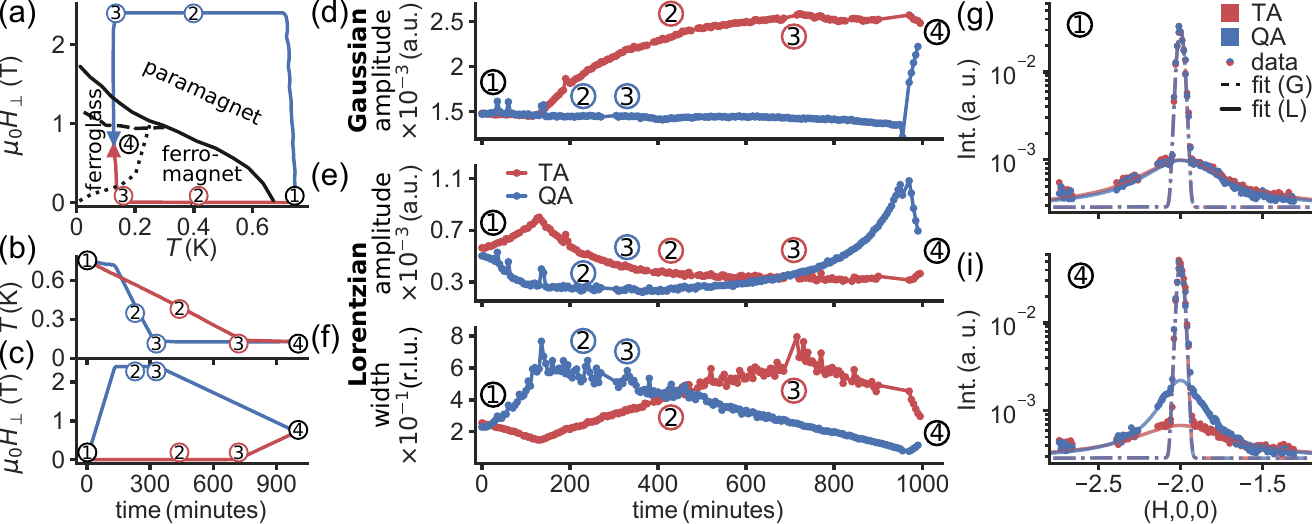}
	\caption{
	Evolution of quantum (blue) versus thermal (red) annealing protocols in \ce{LiHo_{x}Y_{1-x}F4} ($x\,=\,0.45$).
	(a) Transverse field-temperature phase diagram of \ce{LiHo_{x}Y_{1-x}F4} ($x\,=\,0.44$) showing the para- and ferromagnetic phases, as well as the ferroglass phase at low temperatures and finite fields. The phase diagram is taken from \reftaken\cite{1997_Bitko_PhD,1999_Brooke_Science}. See main text for details on how the phase diagram was obtained. 
	(b) Temperature and (c) field as a function of time for the quantum and thermal annealing protocols shown in (c).
	(d-f) Fit results for fitting intensity vs. $H$ cuts through the $(-2,0,0)$-Bragg peak (see \supp\ref{sup:linecut_fitting} for details). The data can be described by a Gaussian (d) plus Lorentzian (e,f) lineshape, corresponding to the Bragg peak and the diffuse scattering, respectively.
	(g,i) Comparison of intensity vs. $H$ cuts during the quantum and thermal annealing at the locations marked in (a-f) as \textbf{1} and \textbf{4}.
	(g) Location \textbf{1} is the start-point of both protocols at \SI{0.75}{K} and zero-field.
	(i) Location \textbf{4} is at the end-point of both protocols at \SI{0.13}{K} and \SI{0.72}{T} in the ferroglass state.
	}
	\label{fig:fig2_pd_prots_lcuts}
\end{figure*}

We now turn to the results of quantum annealing in disordered \ce{LiHo_{x}Y_{1-x}F4}. We performed careful annealing protocols designed to follow exactly the protocols used by \reftaken\cite{1999_Brooke_Science}. The protocols as a function of field, temperature, and time are shown in \fig\ref{fig:fig2_pd_prots_lcuts}(a-c). To obtain a high density of data points during the protocols, scans of the $(-2,0,0)$ peak were collected every \SI{5}{min}. Due to the efficient design of CORELLI, these complete scans across the peak do not involve any moving parts; the sample remained fixed and we simply started a new dataset every 5 minutes.  In contrast, to cover a large volume of reciprocal space, as shown in \fig\ref{fig:fig1_pd_lcuts_3d}(d), the sample must be rotated.  This ``full scan'' takes more than \SI{1.5}{h} to collect, and therefore these were only done at the start and the end of the protocols.

Considering a cut along $H$, through the center of the pinch-point, the observed intensity is well-described by a Gaussian plus Lorentzian lineshape, corresponding to the Bragg scattering and diffuse scattering, respectively. Thus, the evolution of the order parameter can be tracked via the Gaussian, and the evolution of the diffuse scattering via the Lorentzian contribution to the lineshape. 

\fig\ref{fig:fig2_pd_prots_lcuts}(d-f) show the evolution of the Bragg peak and diffuse intensity as a function of time for the quantum and thermal annealing protocols. The function used for fitting the experimental data is described by
\begin{align}
	\label{eq:linecut}
     I(\vec{q}) \,=\, 
        \underbrace{\frac{A_{\mathrm{G}}}{\sqrt{2\pi\sigma^2_{\mathrm{G}}}}\exp{\left\{-\frac{q_H^2}{2\sigma_{\mathrm{G}}^2}\right\}} }_{\mathrm{Gaussian}}
        + 
        \underbrace{\frac{A_{\mathrm{L}}}{\pi}\left( \frac{\gamma_{\mathrm{L}}}{q_H^2 + \gamma_{\mathrm{L}}^2} \right) }_{\mathrm{Lorentzian}}
        +
        I_0
\end{align}
with the wave-vector along the $H$ direction relative to the zone center, $q_H\,=\,Q_H+2$, amplitude $A$, 
standard deviation $\sigma$ of the Gaussian (G) and half width at half maximum $\gamma$ of the Lorentzian (L), respectively, and a constant background $I_0$. 
The standard deviation of the Gaussian $\sigma_{\mathrm{G}}$ was fixed to the resolution of the instrument, \cf \supp\ref{sup:instrumental_resolution}. Each datapoint in \fig\ref{fig:fig2_pd_prots_lcuts}(d-f) was obtained by fitting \eq{\ref{eq:linecut}} to an intensity vs. $H$ cut through the $(-2,0,0)$ zone center collected during a 5 minute window of the protocols, as shown for a single fit in the example in the \supp\ref{sup:linecut_fitting}, \fig\ref{fig:fig_sup_linecut_mesh_example}. During these 5 minutes, the field changes by \SI{0.0125}{T} for QA and the temperature changes by \SI{0.005}{K} fot TA.

We first consider the evolution of the scattering during the QA (blue) protocol. The Gaussian amplitude in \fig\ref{fig:fig2_pd_prots_lcuts}(d) stays constant in the paramagnetic phase. There is a sharp increase in Bragg intensity when entering the ferroglass, consistent with the development of ferromagnetic long range order. The transition into the ferroglass is also accompanied by a sharp maximum in the amplitude of the Lorentzian and a sharp minimum in its width, \cf \fig\ref{fig:fig2_pd_prots_lcuts}(b,c). The protocol ends with a significant Lorentzian amplitude, and thus significant diffuse magnetic scattering.

The evolution of the scattering during the TA protocol is shown in red. The transition into the ferromagnet at $T_{\mathrm{C}}$ shows an increase in Bragg intensity that follows the expected behavior for entering a ferromagnet. The transition into the ferromagnet is also accompanied by a sharp maximum in the amplitude of the Lorentzian and a sharp minimum in its width, \cf \fig\ref{fig:fig2_pd_prots_lcuts}(b,c). This is expected for the critical scattering at a second order phase transition. The protocol ends with diffuse intensity, but much less as compared to the QA protocol.

The amount of diffuse magnetic scattering at the identical location in the phase diagram---the end-point at \SI{0.13}{K} and \SI{0.72}{T} of each protocol with the same overall runtime---is different depending on the annealing protocol. This becomes even more apparent when directly comparing intensity vs. $H$ cuts, shown in \fig\ref{fig:fig2_pd_prots_lcuts}(g,i). These compare QA to TA at the start-point and end-point of the protocols, indicated by the numbers \textbf{1} and \textbf{4} in \fig\ref{fig:fig2_pd_prots_lcuts}(a-c). Both annealing protocols start with identical diffuse and Bragg scattering shown in \fig\ref{fig:fig2_pd_prots_lcuts}(g). At the end-point of both protocols---in the ferroglass at \SI{0.13}{K} and \SI{0.72}{T}---the scans in \fig\ref{fig:fig2_pd_prots_lcuts}(j) showing TA and QA for the first time at the same temperature and field point in the phase diagram since starting the protocols. Despite being at the same location, the diffuse scattering, as indicated by the Lorentzian contribution to the signal, shows very different behavior. After the QA protocol, much stronger diffuse intensity is observed. Given that the end-point is near the phase boundary, and strong critical scattering is expected there \cite{2016_Babkevich_PhysRevB}, this suggests that the QA protocol has reached a configuration closer to equilibrium more quickly. This is further supported by the time-dependence of the scattering after the protocols are completed, as discussed next.

\begin{figure*}[htbp]
	\centering
	\includegraphics[width=\sdcol]{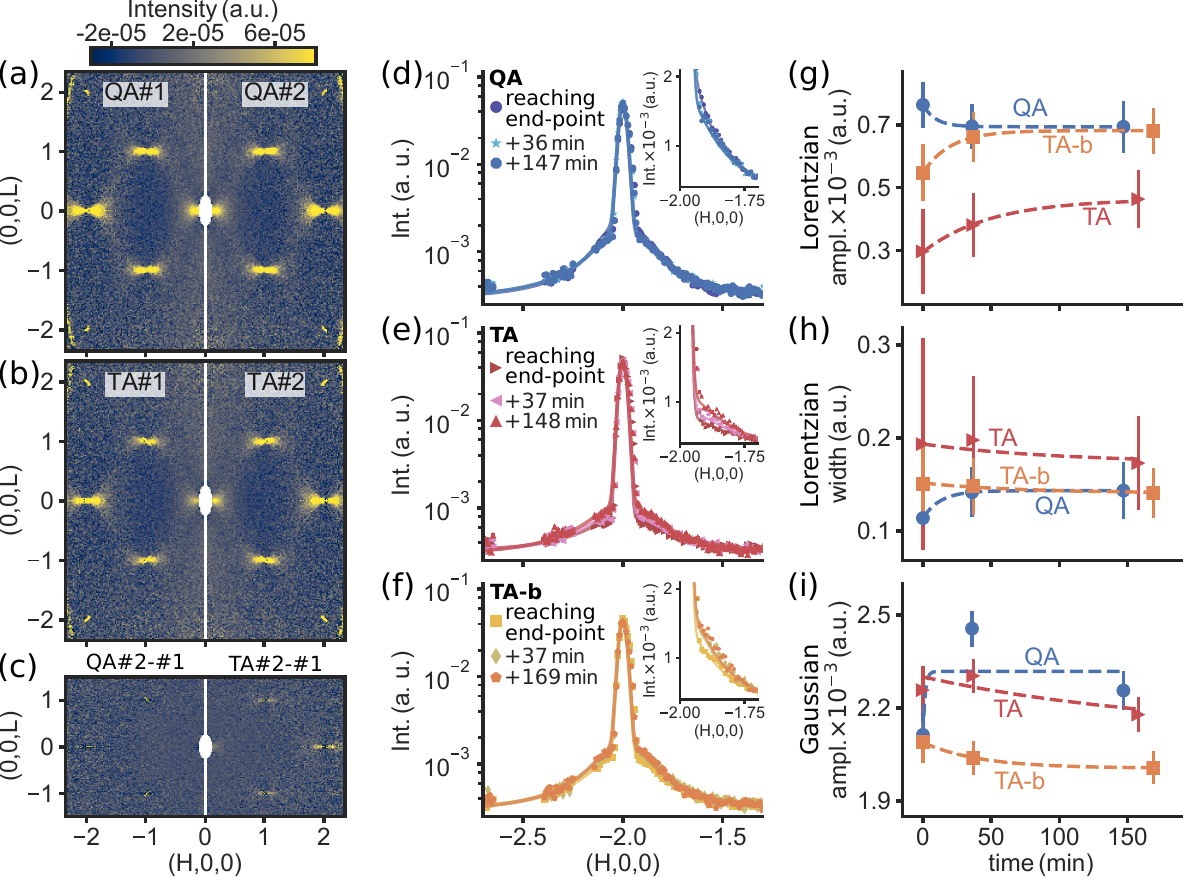}
	\caption{ 
	Diffuse neutron scattering data collected in \ce{LiHo_{x}Y_{1-x}F4} ($x\,=\,0.45$) at the end-point of the annealing protocols.
	(a) Scattering intensity in the $(H,0,L)$-plane ($\mathrm{K}\,=\,[-0.01,0.01]$) directly after the QA protocol, QA\#1 (left panel), and again after $\sim$\SI{2}{h}, QA\#2 (right panel).
	(b) Scattering intensity directly after the TA protocol, TA\#1 (left panel), and again after $\sim$\SI{2}{h}, TA\#2 (right panel).
	(c) Difference of the scattering intensity between the two scans after the QA (left panel) and TA (right panel), respectively. After QA the difference fluctuates about zero since QA\#1 and QA\#2 are identical. After TA there is remaining intensity in the pinch-point scattering indicating that TA\#1 and TA\#2 are not identical.
	(d-f) Semi-log plot of intensity vs. $H$ cuts through the $(-2,0,0)$-Bragg peak (see \supp\ref{sup:linecut_fitting} for details) at the end of the protocols, during the first scan, and during the second scan. Data are shown for QA (d), TA (e), and TA-b (see main text for details) (f). The insets show a zoom-in on a linear scale
	(g-i) Lorentzian amplitude (g) and width (h), and Gaussian amplitude (i) as a function of time after the protocols in (d-f). Data points were obtained by fitting \eq{\ref{eq:linecut}} to the intensity vs. $H$ cuts in (d-f). Broken lines are weighted least square exponential fits, \cf \supp \tab\ref{tab:exp_QAvsTA_endpoint}.
	}
	\label{fig:fig3_QAvsTA_endpoint}
\end{figure*}

The different diffuse scattering at the end-point not only shows much less diffuse intensity after TA compared to QA, but this difference also survives for a long time (longer than our observation time). This is shown in \fig\ref{fig:fig3_QAvsTA_endpoint}, where panels (d-i) show the time evolution of scattering features using three points in time. The first point is the last time step of the protocols. The second and third points are obtained from full volume scans, \cf panels (a,b), each taking about \SI{2}{h} to complete.  We call these QA (or TA) \#1 and \#2. QA\#1 and \#2 are identical, as shown in the difference plots in panel (c), while TA\#1 and \#2 are different. \fig\ref{fig:fig3_QAvsTA_endpoint}(d) and (e) highlight the time evolution of the diffuse scattering by an intensity vs. $H$ cut through the $(-2,0,0)$-Bragg peak at different times after reaching the end-point of the respective annealing protocol. While the scans taken after QA look almost identical, the scans taken after the TA show an increase in diffuse scattering as a function of time. Additionally, \fig\ref{fig:fig3_QAvsTA_endpoint}(f) shows a second thermal annealing protocol (TA-b) that also shows the time-dependent behavior of the initial TA, again stronger than that of QA. This second thermal annealing protocol TA-b took a total time of \SI{7.5}{h} (in comparison to the \SI{16}{h} of QA and TA) by following the same path as TA with faster cooling but the same field ramp rate.  \fig\ref{fig:fig3_QAvsTA_endpoint}(g-h) show the evolution of the Lorentzian amplitude ($A_L$), \ie a measure for the ``amount'' of diffuse intensity, Lorentzian width ($\gamma_L$), and Gaussian amplitude ($A_G$) as a function of time for the different QA and TA protocols. It seems that, after either TA protocol, the system ends with less diffuse scattering and then slowly evolves towards a final state with more diffuse scattering. Assuming an exponential relaxation, TA evolves to a final state with much less diffuse scattering than the final state reached by QA. TA-b, on the other hand, starts with less diffuse scattering than QA and then evolves rather quickly (in comparison to the relaxation time of TA) towards the final state reached by QA. Due to the much faster protocol for TA-b (it takes only about half of the time of either QA and TA), the system is spending less time in the regime with slow dynamics and, thus, it is difficult to directly compare TA-b to the other two protocols. Nonetheless, TA-b is an important measurement as it shows that the path along TA (and TA-b) produces a state with less diffuse scattering that evolves towards a state with more diffuse scattering, as the one QA has reached. The small changes in the Bragg intensity \fig\ref{fig:fig3_QAvsTA_endpoint}(i) are consistent with the trends seen in the diffuse scattering; they are anti-correlated, indicating the trade-off between long- and short-range order, as expected based on the sum rule for magnetic neutron scattering.  

We note that the ac susceptibility results of Brooke \etal are similar to ours; in those data, the ac susceptibility of the TA and QA final states never converge, even after 36 hours of observation. Although we have limited time-dependence of the diffuse scattering data following the protocols, it appears that our results are consistent with this result. To robustly determine the relaxation behavior of the scattering signatures and test the correspondence between these results further, it would be necessary to measure the diffuse scattering in faster succession and for longer times after the protocols are finished. This could be efficiently accomplished using the same setup we have described here, but continuing the short scans after the quenches rather than performing full volume scans.  

There is one consideration which could affect the conclusions outlined above, which is related to beam-heating of the sample. During the experiments the neutron beam was off for some periods of time due to accelerator issues at the spallation source. In particular, during TA the beam was off for \SI{70}{min} towards the end of the protocol and came back on \SI{30}{min} before the end. During QA, the beam was off for \SI{15}{min} towards the end of the protocol and came back on \SI{30}{min} before the end. In both cases the temperature with the beam off reached a base temperature of \SI{0.125}{K}, and in both cases the temperature increased back to \SI{0.128}{K} within $\sim$\SI{10}{min} after the beam returned. Even though the recorded temperature changes only by a few milli-Kelvin and reacts very quickly, this temperature is measured at the cold finger of the dilution refrigerator, rather than at the sample, which is about a \SI{10}{cm} distance away and connected by a copper rod and the copper sample holder.  Thus, the temperature of the sample might show a more dramatic heating/cooling behavior that isn't detected by the thermometer at the cold-finger. For details on the temperature monitoring during the experiment see the \supp\ref{sup:sample_and_sample_orientation}. In case of the second thermal annealing, TA-b, the beam was on for the entirety of the protocol, and thus TA-b should be unaffected by the beam on/off behavior.

With these aspects considered, there  is a clear trend for TA compared to QA. TA takes longer to achieve an equilibrium (or at least, approximately stationary) state, while QA reaches such a state more quickly. Additionally, TA does not reach the same state as QA after 3 hours, and an exponential fit to the data suggests it may never reach it. This strongly suggests that QA is faster at optimizing the spin configurations.

\subsection{Random field effects} 
\label{sub:random_field_effects}

\begin{figure}[htbp]
	\centering
	\includegraphics[width=2.25in]{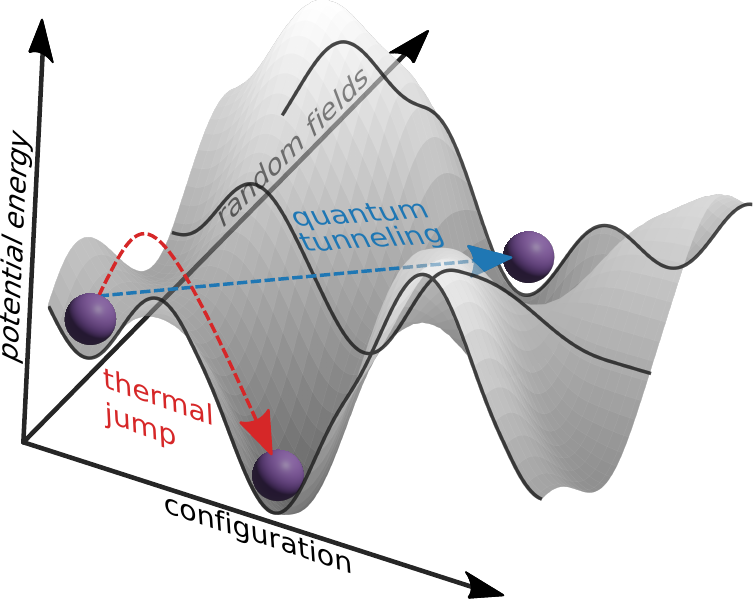}
	\caption{
	Energy landscape in the randomly disordered dipolar coupled Ising ferromagnet in a transverse field. The energy landscape changes as a function of random fields induced by an applied transverse field, thus changing the optimization problem as QA optimization progresses.
	}
	\label{fig:fig4_QA_rdmfields}
\end{figure}

The ordinary conception of QA involves an energy landscape that is fixed throughout the annealing process.  However, as discussed in the introduction, it is clear from prior theoretical and experimental works that when a transverse field is applied to \ce{LiHo_{x}Y_{1-x}F4}, random longitudinal fields are developed. These create a new source of randomness aside from the effective exchange randomness produced by the dilution.  The behavior can be modelled as
\begin{align}
\mathcal{H} \,=\, \sum_{(ij)} J_{ij} \epsilon_i \epsilon_j S£_i^z S_j^z - H_\perp \sum_i \epsilon_i S_i^x - \sum_i H_{||i} \epsilon_i S_i^z,
\end{align}
where $\epsilon_i \,=\, 1$ if a site is occupied by \ce{Ho}, and 0 if it is occupied by \ce{Y}, and $H_{\parallel i}$ is a random field which depends on $H_\perp$.  Not only does this additional random field mean that the behavior should deviate from MFT (which has been the main topic of consideration in the literature so far), but importantly, it should also dramatically influence the quantum annealing process. This is illustrated in the schematic in \fig\ref{fig:fig4_QA_rdmfields}, which shows the scenario of a changing energy landscape during quantum and thermal annealing. The connection between these two aspects deserves further study.  It is thus important to characterize how the spin correlations deviate from MFT expectations in the presence of a transverse field, since the spin correlations can be thought of as a consequence of the annealing process in the presence of the transverse and random fields.

We have observed the deviation from MFT of our diffuse scattering data, which directly reflects the microscopic spin correlations. In the following, we compare the diffuse magnetic neutron scattering in zero-field---and thus without random fields---to data obtained in an applied magnetic field where random fields should be present.

\begin{figure*}[htbp]
	\centering
	\includegraphics[width=\sdcol]{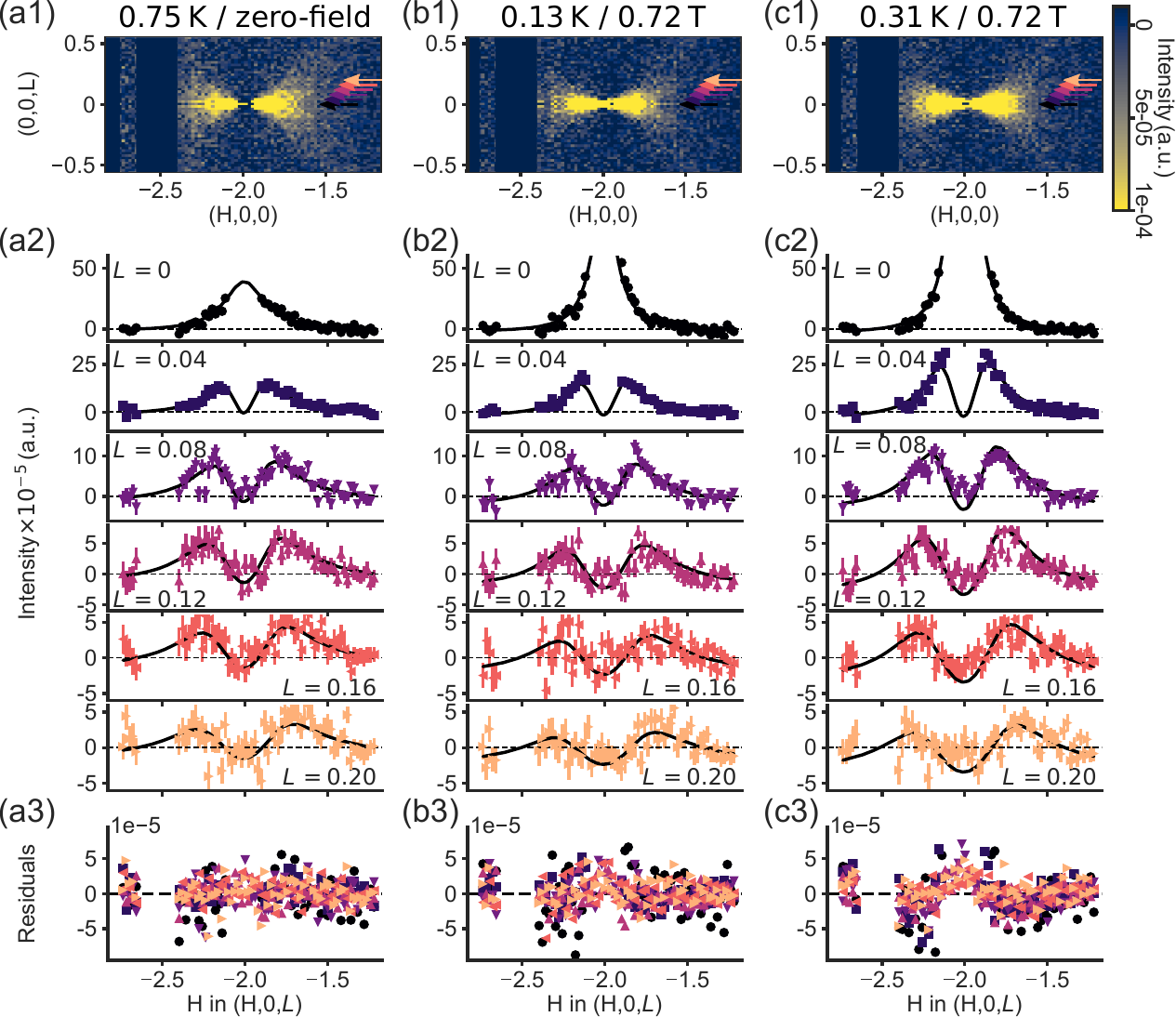}
	\caption{
	Pinch-point scattering in the vicinity of the $(-2,0,0)$ magnetic Bragg peak.
	Data are shown in zero-field at \SI{0.75}{K} (a) and in an applied transverse field of \SI{0.72}{T} at \SI{0.13}{K} (b) and \SI{0.31}{K} (c).
	(a1-c1) Scattering intensity in the $(H,0,L)$-plane ($K\,=\,[-0.01,0.01]$). Regions with intensity from powder scattering from the sample holder are masked.
	(a2-c2) Cuts through the pinch-point along $H$ for different $L$-values ($\delta L\,=\,0.04$ and $K\,=\,[-0.01,0.01]$) as indicated by the arrows in (a1-c1). Solid lines are fits to the data using \eq{\ref{eq:alsnielsen}}, with the fit results summarized in \tab\ref{tab:pinchpoint_fits}. As mentioned in the main text, data were fitted by constraining the transverse correlation length.
	(a3-c3) Residuals of the fits in (a2-c2).
	}
	\label{fig:fig5_pinchpoint_fits}
\end{figure*}

\fig\ref{fig:fig5_pinchpoint_fits} shows the pinch-point scattering in the vicinity of the $(-2,0,0)$ magnetic Bragg peak at different temperatures and transverse fields. Background taken in the field-polarized state at \SI{0.13}{K} in a \SI{3}{T} field was subtracted from the data. The data compared here were taken at (a) \SI{0.75}{K} in zero-field in the paramagnet just above the phase transition, (b) \SI{0.13}{K} in \SI{0.72}{T} in the ferroglass just below the phase transition, and (c) \SI{0.31}{K} in \SI{0.72}{T} in the ferromagnet just below the phase transition. Panels (a1-c1) show the pinch-point scattering intensity in the $(H,0,L)$-plane surrounding the $(-2,0,0)$-Bragg peak position. Intensity vs. $H$ cuts through the pinch-points are shown in panels (a2-c2). The solid lines are fits to the experimental data using the magnetic neutron scattering cross section of Ising moments in the quasielastic limit convoluted with the instrumental resolution given by
\begin{align}
	\label{eq:convofit}
    \left(\frac{\mathrm{d}\sigma}{\mathrm{d}\Omega} \circledast R \right)(\vec{Q}) \,=\, \int_{\mathbb{D}} \frac{\mathrm{d}\sigma}{\mathrm{d}\Omega}(\vec{Q^{\prime}}) R(\vec{Q}-\vec{Q^{\prime}})\mathrm{d}\vec{Q^{\prime}}.
\end{align}
The instrumental resolution is represented by a Gaussian distribution (\cf \supp\ref{sup:instrumental_resolution}) and the elastic neutron scattering cross section is related to the wave-vector dependent susceptibility, $\chi(\vec{Q})$, as follows:
\begin{align}
    \label{eq:crosssec}
    \frac{\mathrm{d}\sigma}{\mathrm{d}\Omega} \,\propto\, \left( 1-\left( \frac{Q_z}{Q} \right)^2 \right) f^2(\vec{Q}) \chi(\vec{Q})
\end{align}
with the wave-vector transfer $\vec{Q}$ (and its component along the Ising direction is denoted $Q_z$), the magnetic form factor $f(\vec{Q})$ for \ce{Ho^{3+}}. The mean-field expression for the wave-vector dependent susceptibility of the dipolar coupled Ising ferromagnet is
\begin{align}
	\label{eq:alsnielsen}
    \frac{1}{\chi(\vec{Q})} \,\propto\, 1+ \xi_{\perp}^{2} \left( q^2 + g \left( \frac{q_z}{q} \right)^2 \right) - h q_z^2,
\end{align}
with a transverse correlation length $\xi_{\perp}$, an anisotropy factor $g$, and the wave-vector $\vec{q}\,=\,\vec{Q}-\vec{\tau}$ with the wave-vector transfer $\vec{Q}$ and the wave-vector of the magnetic Bragg peak at $\vec{\tau}$. Thereby, $q_z$ is along the Ising direction ($L$ in our case) and $q$ perpendicular to it ($H$ in the case of the data shown in \fig\ref{fig:fig5_pinchpoint_fits}). As the parameter $h$ is zero at $T_{\mathrm{C}}$, the term $h q_z^2$ will be neglected for analyzing the data in \fig\ref{fig:fig5_pinchpoint_fits} since all data sets were taken close to the phase boundary \cite{1976_Als-Nielsen_PhysRevLett}. From $\xi_{\perp}$ and $g$ also follows a longitudinal correlation length $\xi_{\parallel}\,=\,\sqrt{g}\xi_{\perp}^2$ \cite{1976_Als-Nielsen_PhysRevLett}. For a detailed derivation of \eq{\ref{eq:crosssec}} and \eq{\ref{eq:alsnielsen}} we refer to \reftaken\cite{1984_Lovesey_,1984_Lovesey_a} and \reftaken\cite{1973_Aharony_PhysRevB,1976_Als-Nielsen_PhysRevLett}, respectively. Data were fitted by constraining the transverse correlation length $\xi_{\perp}$ to the result obtained from fitting the intensity vs. $H$ cut at $L\,=\,0$ of the data without background subtraction to include the information of the Bragg peak. This is a necessary step since background subtracted data misses crucial information in the zone center that is important for obtaining physical meaningful fit results. For details on the fitting procedure see \supp\ref{sup:pinch_point_fitting}. Panels (a3-c3) show the residuals of the fits and \tab\ref{tab:pinchpoint_fits} summarizes the obtained fit parameters.

\begin{table}[htbp]
\centering
\caption{Results of fitting the mean-field expression of the wave-vector dependent susceptibility in \eq{\ref{eq:convofit}} to the data in \fig\ref{fig:fig5_pinchpoint_fits}. The quality of the fit is represented by $\chi_{\nu}^2$ reduced by the degrees of freedom $\nu$, and the residual sum of squares (RSS). $^{*}$ these correlation lengths are at the upper bound of the constraints, thus the uncertainty is asymmetric.}
\label{tab:pinchpoint_fits}
\begin{tabularx}{\columnwidth}{lccccc}
\hline\hline
             & $\xi_{\perp}\,(\mathrm{\AA})$                      & $\xi_{\parallel}\,(\mathrm{\AA})$                   & $g\,(\mathrm{\AA}^{-2})$       &  $\chi_{\nu}^{2}$           & RSS                       \\
             \hline
\SI{0.75}{K} & \multirow{2}{*}{$5.7(4)$}                       & \multirow{2}{*}{$34(5)$}                       & \multirow{2}{*}{$1.1(1)$} & \multirow{2}{*}{1.18} & \multirow{2}{*}{$1.2$e-7} \\
zero-field   &                                                    &                                                     &                                &                      &                           \\ 
\SI{0.13}{K} & \multirow{2}{*}{$9.8(3)^{*}$}                     & \multirow{2}{*}{$112(8)$}  & \multirow{2}{*}{$1.4(1)$} & \multirow{2}{*}{1.51} &  \multirow{2}{*}{$1.8$e-7} \\
\SI{0.72}{T} &                                                    &                                                     &                                &                      &                           \\ 
\SI{0.31}{K} & \multirow{2}{*}{$16(2)^{*}$} & \multirow{2}{*}{$3(1)$e2} & \multirow{2}{*}{$1.5(1)$} & \multirow{2}{*}{2.02} & \multirow{2}{*}{$3.4$e-7} \\
\SI{0.72}{T} &                                                    &                                                     &                                &                      &   
\\\hline\hline
\end{tabularx}
\end{table}

On first glance, the mean-field expression for the wave-vector dependent susceptibility seems to describe our data well for all three locations in the phase diagram. However, the agreement is notably better for the data taken in zero-field, shown in \fig\ref{fig:fig5_pinchpoint_fits}(a). In zero-field, the model in \eq{\ref{eq:alsnielsen}} represents the data and the fit produces physically meaningful results. In an applied transverse field, \cf \fig\ref{fig:fig5_pinchpoint_fits}(b) and (c), the fit is getting worse, especially at small $L$ close to the center of the pinch-point, as can be also seen in the residuals and $\chi^2$ (see also \tab\ref{tab:pinchpoint_fits}). This is also due to the necessary constraining of the fit as discussed in \supp\ref{sup:pinch_point_fitting}. Data in zero-field delivers the same result fitting with and without constraints, converging on an excellent fit for all $L$. This is in contrast to data taken in an applied transverse field, where constraining the fit to physically meaningful values noticeably reduces the fit quality, which speaks to the inappropriateness of the model for these in-field data.  In particular, the model underestimates the intensity near $H=0$, and overestimates it away from $H=0$, for all $L$. Thus, in zero-field, the data can be described within mean-field theory via \eq{\ref{eq:alsnielsen}}, which is not the case for data taken in an applied transverse field, indicative of the influence of random fields \cite{2007_Silevitch_Nature}.

The agreement of \eq{\ref{eq:alsnielsen}} with the zero-field data over the whole measured range is illustrated by the mean-field simulation in \fig\ref{fig:fig1_pd_lcuts_3d}(e). The simulated data were obtained by first fitting the mean-field expression to the experimental data taken close to the zone center at \SI{0.75}{K} in zero-field (\fig\ref{fig:fig5_pinchpoint_fits}(a)). From these fit results the large volume in reciprocal space in \fig\ref{fig:fig1_pd_lcuts_3d} was calculated. It should be mentioned here that the mean-field form we have used is only expected to be quantiatively valid close to the Bragg peak, however we find that it still qualitatively reproduces features throughout the whole Brillouin zone, and hence reproduces the whole volume of data well.   This shows that all of the diffuse scattering in zero-field can be well described as the effects of dipolar coupled Ising moments.

\section{Conclusion}
\label{sec:conclusion}

We studied the microscopic spin correlations in the disordered transverse field dipolar Ising model magnet \ce{LiHo_{x}Y_{1-x}F4} using diffuse magnetic neutron scattering. We successfully tested the effects of quantum vs. thermal annealing in the ferroglass state, and demonstrated that there is a general trend where the diffuse scattering after quantum annealing is almost static while, after thermal annealing, it evolves more slowly with time. This suggests that quantum annealing does enable the system to reach a state closer to equilibrium more quickly.  Additionally, through a detailed analysis of large volume scans in reciprocal space we confirm the departure from mean field theory when a transverse field is applied. This departure is thought to be caused by random fields that are induced by the transverse field in the diluted dipolar coupled Ising system.  These external-field-dependent random fields will change the energy landscape of the system as the quantum annealing progresses. Thus, not only quantum fluctuations are added to the system to perform the optimization, but the optimization problem itself changes. This is in contrast to the ``simple'' quantum tunneling picture that is studied in most ideas of QA for quantum computations, as well as was initially considered by Brooke \etal \cite{1999_Brooke_Science} for \ce{LiHo_{x}Y_{1-x}F4}, where the Hamiltonian of the optimization problem stays unchanged and only quantum fluctuations are added to the system.  We have shown that \ce{LiHo_{x}Y_{1-x}F4} in a transverse field is not a realization of this simple case, instead realizing a more complex QA picture, yet this more complex QA procedure still appears to offer a quantum speedup.  This type of ``changing landscape'' QA may indeed be worth studying in the context of quantum information processing.

\begin{acknowledgements}
The authors wish to thank M.J.P. Gingras, D. Lozano-Gómez, and J.A.M. Paddison for support and stimulating discussions.
This research was funded by the U.S. Department of Energy, Office of Science, Basic Energy Sciences, Materials Sciences and Engineering Division under Award Number DE-SC0018972.
This research was partially supported by the CIFAR Quantum Materials program.
This research used resources at the Spallation Neutron Source, a DOE Office of Science User Facility operated by the Oak Ridge National Laboratory.
\end{acknowledgements}


\section*{Supplemental Materials}

In the supplemental material we describe the experimental setup and discuss the orientation of the sample with respect to the magnetic field and the scattering plane. We also describe the data reduction process for fitting the intensity vs. $H$ cuts in \sect\ref{sub:quantum_vs_thermal_annealing} and the pinch-point fitting in \sect\ref{sub:random_field_effects}, and discuss the instrumental resolution.

\subsection{Sample and Sample Orientation}
\label{sup:sample_and_sample_orientation}

\begin{figure}[htbp]
	\centering
	\includegraphics[width=\scol]{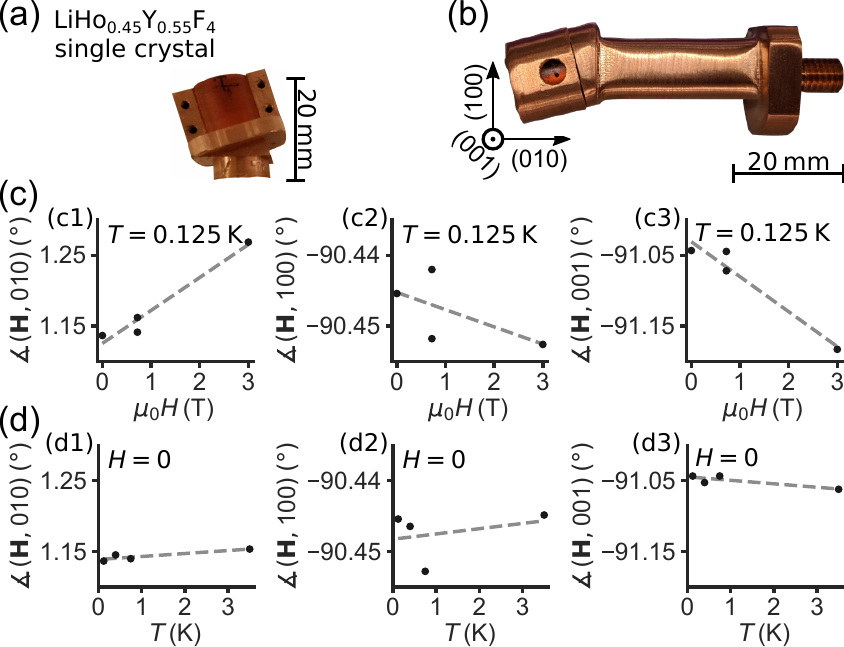}
	\caption{Sample and sample alignment for the neutron scattering experiments.
	(a) High-quality single crystal \ce{LiHo_{x}Y_{1-x}F4} ($x\,=\,0.45$) with a total mass of \SI{3.82}{g}. The cylindrical shape of the sample has a height of \SI{9.4}{mm} and a diameter of \SI{10.6}{mm}. The crystallographic $a$-axis (or $b$-axis) is tilted \SI{14.01}{\degree} from the cylinder axis, thus the field is applied with \SI{14.01}{\degree} tilt along the cylinder axis.
	(b) Sample in the oxygen-free copper sample holder composed of a body and lid accurately matching the dimensions of the sample. The sample holder has a total mass of \SI{46.2}{g}.  Field was applied along the length of the sample holder, the $(010)$-direction of the crystal. The scattering plane is spanned by the $(100)$ and $(001)$-directions.
	(c,d) Sample misalignment as a function of field at \SI{0.13}{K} (c1-c3) and temperature at zero-field (d1-d3), respectively. The plots show the angle between the applied magnetic field, \ie out-of-scattering-plane, and the three main symmetry directions $(010)$, $(100)$, and $(001)$.
	}
	\label{fig:fig_sup_samplealignment}
\end{figure}

The sample and sample holder used for the neutron scattering experiments are shown in \fig\ref{fig:fig_sup_samplealignment}(a) and (b), respectively. The high-quality single crystalline sample of \ce{LiHo_{x}Y_{1-x}F4} ($x\,=\,0.45$) is of cylindrical shape with a height of \SI{9.4}{mm}, a diameter of \SI{10.6}{mm}, and a total mass of \SI{3.82}{g}. The crystallographic $a$-axis (or $b$-axis) is tilted by \SI{14.01}{\degree} with respect to the cylinder axis. The oxygen-free sample holder accurately matches the shape of the sample, thus providing excellent thermal contact. A lid allows rigidly fixing the position of the sample against torques from the applied field. Additionally, the head of the sample holder containing the sample is tilted by \SI{14.01}{\degree} so that the crystallographic $a$-axis (or $b$-axis) lies along the length of the sample holder. The sample holder was firmly bolted into the copper cold-finger of the mixing chamber of the dilution unit. The dilution unit was inserted into the vertical \SI{5}{T} SLIM-SAM magnet. Sample temperature was monitored with a ruthenium oxide sensor mounted at the end of the cold-finger very close to the sample holder.

During our experiment we defined the $b$-axis ($(0,1,0)$) to lie along the cylindrical axis of the sample holder, thus the $a$- ($(1,0,0)$) and $c$- ($(0,0,1)$) axes span the scattering plane. Hence, the applied magnetic field is along the $(0,1,0)$-direction and perpendicular to the $(1,0,0)$ and---most importantly---to the Ising direction $(0,0,1)$. \fig\ref{fig:fig_sup_samplealignment}(c,d) show the angle between the applied magnetic field and the three main symmetry directions as a function of field (c) and temperature (d), indicating a transverse field misalignment with respect to the Ising direction of about \SI{1.2}{\degree}.

\subsection{Data Reduction}
\label{sup:data_reduction}

Initial data reduction of the neutron scattering data obtained at CORELLI was performed using MANTID \cite{2013_Akeroyd_Software,2014_Arnold_NIM}. The initial data reduction includes: determining the UB matrix; translating angular data into $(H,K,L)$-space; binning the raw data, data showing $(H,0,L)$-planes are binned with step sizes: $\delta H\,=\,\delta K\,=\,\delta L\,=\,0.02\,$r.l.u., intensity vs. $H$ cuts for quantum vs. thermal annealing were binned with $\delta H\,=\,0.0083\,$r.l.u. and $\delta K\,=\,\delta L\,=\,0.1\,$r.l.u., intensity vs. $H$ cuts for pinch-point fitting were binned with $\delta H\,=\,0.0083\,$r.l.u. and $\delta K\,=\,\delta L\,=\,0.04\,$r.l.u.; data normalization.

\subsubsection{Intensity vs. $H$ Cut Fitting}
\label{sup:linecut_fitting}

\begin{figure}[t]
	\centering
	\includegraphics[width=\scol]{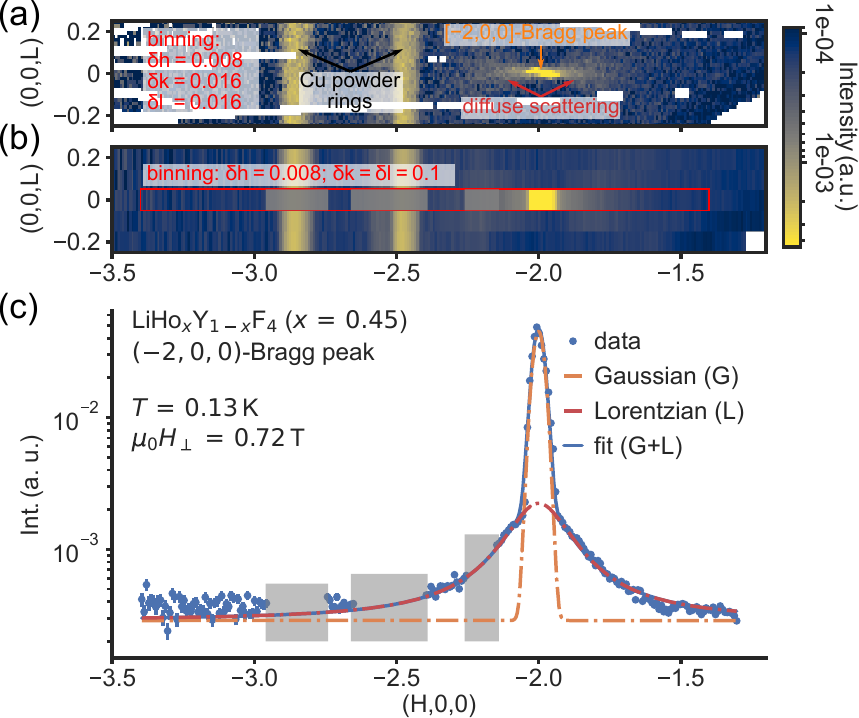}
	\caption{Intensity vs. $H$ cut data reduction for the protocol measurements in \ce{LiHo_{x}Y_{1-x}F4} ($x\,=\,0.45$). Data shown were taken at the end-point of the quantum annealing protocol at (\SI{0.13}{K}; \SI{0.72}{T}).
	(a,b) Data in the $(H,0,L)$-plane obtained from a single scan at fixed sample rotation focused on the $(-2,0,0)$-Bragg peak. The data show the Bragg peak and the diffuse scattering emerging from it. The region of interest shows three copper powder rings from the sample holder (the ring at $H\,\sim\,2.2\,$r.l.u. is barely visible).
	(b) The same data as in (a), plotted with the binning used for the intensity vs. $H$ cut extraction, \ie $\delta H\,=\,0.008$; $\delta K\,=\,0.1$; $\delta L\,=\,0.1$. The red box indicates the intensity vs. $H$ cut, and the gray boxes the masked region of copper powder ring intensity.
	(c) Semi-log plot of an intensity vs. $H$ cut through the $(-2,0,0)$-Bragg peak. The experimental data is well described by a Gaussian (broken orange line) plus Lorentzian (broken red line) lineshape, describing the Bragg peak and diffuse scattering, respectively. The solid blue line is a fit to the data using \eq{\ref{eq:linecut}}. The gray boxes indicate the position of powder rings and were cut out of the data prior to fitting.}
	\label{fig:fig_sup_linecut_mesh_example}
\end{figure}

\fig\ref{fig:fig_sup_linecut_mesh_example} describes the data reduction performed for all intensity vs. $H$ cuts obtained during the QA vs. TA analysis, \ie data shown in \fig\ref{fig:fig2_pd_prots_lcuts} and \fig\ref{fig:fig3_QAvsTA_endpoint}. \fig\ref{fig:fig_sup_linecut_mesh_example}(a)	shows the $(H,0,L)$-plane surrounding the $(-2,0,0)$-Bragg peak obtained for a single sample rotation. The binning for the cut extraction of the same data is shown in \fig\ref{fig:fig_sup_linecut_mesh_example}(b). In \fig\ref{fig:fig_sup_linecut_mesh_example}(c) the intensity of the intensity vs. $H$ cuts through the $(-2,0,0)$-Bragg peak is plotted as a function of $(H,0,0)$. The experimental data is described by a Gaussian plus Lorentzian lineshape, describing the Bragg peak and diffuse scattering, respectively. Therefore, the function used for fitting is
\begin{align}
	\label{eq-sup:linecut}
     &f(x;A_{\mathrm{G}},\mu_{\mathrm{G}},\sigma_{\mathrm{G}},A_{\mathrm{L}},\mu_{\mathrm{L}},\sigma_{\mathrm{L}},m,b) \,= \quad\quad\quad \nonumber\\
     &\quad \frac{A_{\mathrm{G}}}{\sigma\sqrt{2\pi}}\exp{\left\{-\frac{(x-\mu_{\mathrm{G}})^2}{2\sigma_{\mathrm{G}}^2}\right\}} + \frac{A_{\mathrm{L}}}{\pi}\left( \frac{\gamma_{\mathrm{L}}}{(x-\mu_{\mathrm{L}})^2 + \gamma_{\mathrm{L}}^2} \right) \nonumber\\
     &\quad+ mx+b,
\end{align}
with the amplitude $A$, center $\mu$, and standard deviation $\sigma$ of the Gaussian (G) and width $\gamma$ of the Lorentzian (L), respectively, and a linear background with slope $m$ and intercept $b$. During fitting, the center of the Gaussian and Lorentzian was fixed to the $(-2,0,0)$-Bragg peak, \ie $\mu\,=\,-2$, and the slope of the background was set to $m\,=\,1$, assuming a constant background. The width of the Bragg peak $\sigma_{\mathrm{G}}$ was determined in the paramagnetic phase and kept fixed for fitting the intensity vs. $H$ cuts during the protocols.

\subsubsection{QA vs. TA Endpoint}

The time dependence of the Lorentzian amplitude ($A_{\mathrm{L}}$), Lorentzian half width at half maximum ($\gamma_{\mathrm{L}}$), and Gaussian amplitude ($A_{\mathrm{G}}$) was fitted using a weighted least square exponential fit of the form
\begin{align}
    \label{eq:exp}
    f(t;a,b,y_0)\,=\,(a-y_0)\exp(-bt)+y_0,
\end{align} 
with the relaxation time $\tau\,=\,1/b$ and the relaxation difference from start to end $\delta y\,=\,a-y_0$. The fit results are summarized in \tab\ref{tab:exp_QAvsTA_endpoint}.

\begin{table}[]
\caption{Results of the weighted least square exponential fitting of the time evolution of the Lorentzian amplitude, Lorentzian half width at half maximum, and Gaussian amplitude after QA, TA, and TA-b in \fig\ref{fig:fig3_QAvsTA_endpoint}. Experimental data was fitted using \eq{\ref{eq:exp}}.}
\label{tab:exp_QAvsTA_endpoint}
\begin{tabular}{llllll}
\hline\hline
\multicolumn{6}{c}{Lorentzian amplitude}                             \\
protocol      & $a$      & $b$      & $y_0$    & $\delta y$ & $\tau\,$(min) \\
\hline\\
\textbf{QA}   & $7.6\times10^{-4}$ & $0.11$ & $6.9\times10^{-4}$ & $7.0\times10^{-5}$   & $9.1$    \\
\textbf{TA}   & $3.0\times10^{-4}$ & $0.02$ & $4.6\times10^{-4}$ & $-1.7\times10^{-4}$  & $49.3$   \\
\textbf{TA-b} & $5.5\times10^{-4}$ & $0.05$ & $6.8\times10^{-4}$ & $-1.3\times10^{-4}$  & $20.3$   \\
              &          &          &          &            &        \\
\multicolumn{6}{c}{Lorentzian width}                                 \\
protocol      & $a$      & $b$      & $y_0$    & $\delta y$ & $\tau\,$(min) \\
\hline\\
\textbf{QA}   & $0.11$ & $0.078$ & $0.14$ & $-0.030$  & $12.9$   \\
\textbf{TA}   & $0.19$ & $0.010$ & $0.17$ & $0.020$   & $101.9$  \\
\textbf{TA-b} & $0.15$ & $0.014$ & $0.14$ & $0.011$   & $71.7$   \\
              &          &          &          &            &        \\
\multicolumn{6}{c}{Gaussian amplitude}                               \\
protocol      & $a$      & $b$      & $y_0$    & $\delta y$ & $\tau\,$(min) \\
\hline\\
\textbf{QA}   & $2.1\times10^{-3}$ & $0.62$ & $2.3\times10^{-3}$ & $-2\times10^{-4}$  & $1.6$    \\
\textbf{TA}   & $2.3\times10^{-3}$ & $0.01$ & $2.1\times10^{-3}$ & $2\times10^{-4}$   & $180.2$  \\
\textbf{TA-b} & $2.1\times10^{-3}$ & $0.02$ & $2.0\times10^{-3}$ & $1\times10^{-4}$   & $41.4$  \\
\hline\hline
\end{tabular}
\end{table}


\subsubsection{Pinch-Point Fitting} 
\label{sup:pinch_point_fitting}

\begin{figure}[t]
	\centering
	\includegraphics[width=\scol]{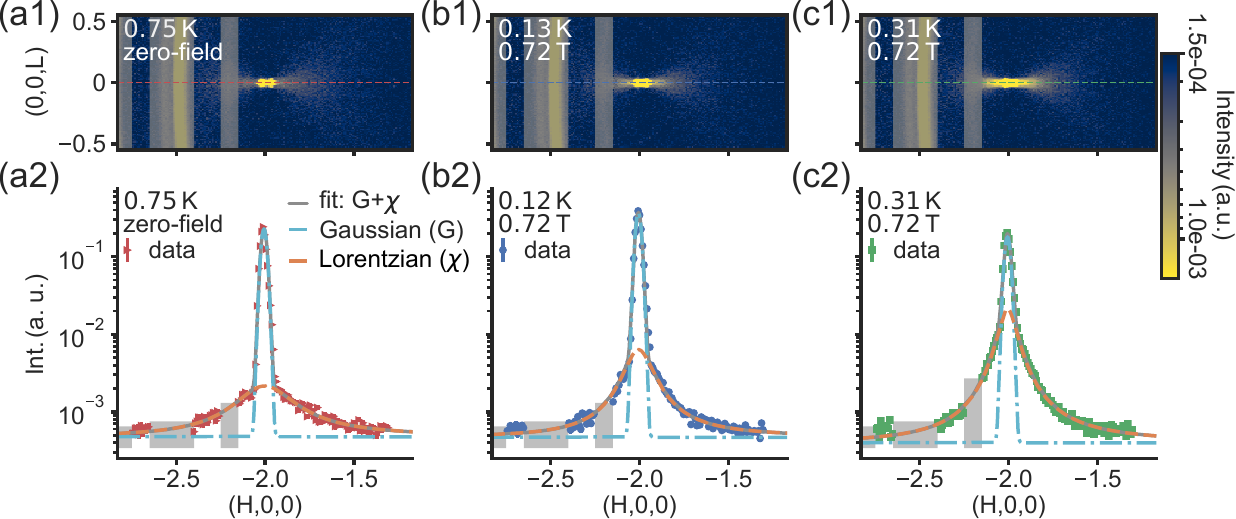}
	\caption{Comparison of fitting the mean-field expression for the wave-vector dependent susceptibility in \eq{\ref{eq:alsnielsen}} with and without including the Bragg peak intensity. (a1-c1) Scattering intensity in the $(H,0,L)$-plane ($K\,=\,[-0.01,0.01]$) at \SI{0.75}{K} in zero-field (a1), \SI{0.13}{K} and \SI{0.72}{T} (b1), and \SI{0.31}{K} and \SI{0.72}{T} (c1). Dashed lines indicate where the data in (a2-c2) was cut. (a2-c2) Cuts through the $(-2,0,0)$-Bragg peak and diffuse scattering along $H$ ($L\,=\,[-0.02,0.02]$ and $K\,=\,[-0.01,0.01]$). Solid gray lines are a fit to the experimental data using a sum of a Gaussian lineshape (representing the Bragg peak) and \eq{\ref{eq:alsnielsen}} with $q_z\,=\,0$. The fit function is separated in the Gaussian (cyan dash-dotted line) and \eq{\ref{eq:alsnielsen}} (orange dashed line). The black dotted line represents a lineshape consisting of the Gaussian from the fit (cyan dash-dotted line) and \eq{\ref{eq:alsnielsen}} with the fit parameters obtained in \fig\ref{fig:fig5_pinchpoint_fits}, \ie including data with $q_z\,\neq\,0$.}
	\label{fig:fig_sup_fixfit}
\end{figure}

The pinch-points in the main text were discussed on background subtracted data. Fitting \eq{\ref{eq:convofit}} to the background subtracted data can lead to not physically meaningful results. The reason for this is that the background subtraction of a strong nuclear and magnetic Bragg peak at the zone center---where the pinch-points emerge from---is not perfect due to high intensity differences, thus data close to the zone center is disregarded. Therefore, when fitting \eq{\ref{eq:convofit}}, crucial data describing the center of the pinch-point is missing, which might lead to a wrong interpretation of the result. To circumvent that, we obtain an upper bound on the correlation length via a fit of the data for $q_z\,=\,0$ \textit{including} the Bragg peak (and thus the background). This is shown in \fig\ref{fig:fig_sup_fixfit} where the experimental data is fitted with a function consisting of a delta-function representing the Bragg peak and the mean-field expression for the wave-vector dependent susceptibility in \eq{\ref{eq:alsnielsen}} at $q_z\,=\,0$ describing the pinch-point scattering, convoluted with a Gaussian representing the instrumental resolution. Using the same approach and fitting the data without background subtraction for $L\,\geq\,0$ is not possible since the background is covering important information further away from the zone center.

Additionally, we can compare the constrained fit in \fig\ref{fig:fig_sup_fixfit} to how the Bragg plus pinch-point scattering would look like without constraining. While for zero-field data the fits obtain the same results, for the data in an applied transverse field, the obtained results for the perpendicular correlation length $\xi_{\perp}$ could not be true values for two reasons: (i) The diffuse scattering would be narrower than the Bragg peak and thus not observable. It becomes apparent that the correlation lengths for fitting the data in \fig\ref{fig:fig5_pinchpoint_fits}(b,c)  need constraining, and upon doing so, the quality of the fit decreases. This demonstrates that the model does not work well for the data in the presence of a transverse field. (ii) In case of the \SI{0.31}{K} data taken in a \SI{0.72}{T} field (\fig\ref{fig:fig5_pinchpoint_fits}(c)) the obtained correlation length would be well above the resolution limit discussed in \supp\ref{sup:instrumental_resolution}.


\subsubsection{Instrumental Resolution}
\label{sup:instrumental_resolution}

The instrumental resolution is closely represented by a Gaussian distribution of the form 
\begin{align}
    I(\vec{q}) \,=\, \frac{A}{\sigma\sqrt{2\pi}} \exp{\left\{ -\frac{(q-\mu)^2}{2\sigma^2} \right\} },
\end{align}
with amplitude $A$ accounting for nuclear and magnetic intensity, variance $\sigma$ describing the width of the resolution, and expected value $\mu$ to correct for any offsets in $q$. The standard deviation $\sigma\,=\,0.0195\,$r.l.u. and expected value $\mu\,=\,-0.0067\,$r.l.u. of the resolution Gaussian were obtained by fitting the field-polarized Bragg peak.

Considering that the correlation length is limited by the half width half maximum (HWHM) of a Gaussian envelope in direct space and is related to the measured full width half maximum (FWHM) Gaussian in reciprocal space by
\begin{align}
	\label{eq:HWHMfwhm}
    \mathrm{HWHM_{\text{direct space}}} &\,=\, \frac{4\ln{2}}{\mathrm{FWHM_{\text{reciprocal space}}}} \nonumber \\
    &\,=\, \frac{\sqrt{2\ln{2}}}{\sigma}.
\end{align}
From the data taken in the field-polarized paramagnetic state at \SI{0.13}{K} and \SI{3}{T}, the correlation length is resolution limited by $\xi_{\mathrm{max}}\,=\,\SI{73}{\AA}$, \ie we could not observe correlation lengths longer than this.

\bibliography{library.bib}

\end{document}